\documentclass[aps,prl,twocolumn,superscriptaddress,amsmath,amssymb,showpacs]{revtex4}

\usepackage{amssymb}
\usepackage[dvips]{graphicx}
\usepackage{bm}
\usepackage{epsfig}
\usepackage[ansinew]{inputenc}

\begin{document}

\title{Co-substitution effects on the Fe-valence in the BaFe$_2$As$_2$ superconducting compound: A study of hard x-ray absorption spectroscopy}

\author{E. M. Bittar}
\email{eduardo.bittar@lnls.br}
\affiliation{Laborat\'{o}rio Nacional de Luz S\'{\i}ncrotron, 13083-970 Campinas, SP, Brazil}

\author{C. Adriano}
\affiliation{Instituto de F\'{\i}sica ``Gleb Wataghin", UNICAMP, 13083-859 Campinas, SP, Brazil}

\author{T. M. Garitezi}
\affiliation{Instituto de F\'{\i}sica ``Gleb Wataghin", UNICAMP, 13083-859 Campinas, SP, Brazil}

\author{P. F. S. Rosa}
\affiliation{Instituto de F\'{\i}sica ``Gleb Wataghin", UNICAMP, 13083-859 Campinas, SP, Brazil}

\author{L. Mendon\c{c}a-Ferreira}
\affiliation{Instituto de F\'{\i}sica e Matem\'{a}tica, Universidade Federal de Pelotas (UFPel), 96010-900 Pelotas, RS, Brazil}

\author{F. Garcia}
\affiliation{Laborat\'{o}rio Nacional de Luz S\'{\i}ncrotron, 13083-970 Campinas, SP, Brazil}

\author{G. de M. Azevedo}
\affiliation{Instituto de F\'{\i}sica, Universidade Federal do Rio Grande do Sul (UFRGS), 91501-970, Porto Alegre, RS, Brazil}

\author{P. G. Pagliuso}
\affiliation{Instituto de F\'{\i}sica ``Gleb Wataghin", UNICAMP, 13083-859 Campinas, SP, Brazil}
\affiliation{Department of Physics and Astronomy, University of California Irvine, 92697-4575, Irvine, CA, USA}

\author{E. Granado}
\affiliation{Instituto de F\'{\i}sica ``Gleb Wataghin", UNICAMP, 13083-859 Campinas, SP, Brazil}
\affiliation{Laborat\'{o}rio Nacional de Luz S\'{\i}ncrotron, 13083-970 Campinas, SP, Brazil}

\date{\today}

\begin{abstract}

The Fe $K$ X-ray absorption near edge structure of BaFe$_{2-x}$Co$_x$As$_{2}$ superconductors was investigated. No appreciable alteration in shape or energy position of this edge was observed with Co substitution. This result provides experimental support to previous \textit{ab initio} calculations in which the extra Co electron is concentrated at the substitute site and do not change the electronic occupation of the Fe ions. Superconductivity may emerge due to bonding modifications induced by the substitute atom that weakens the spin-density-wave ground state by reducing the Fe local moments and/or increasing the elastic energy penalty of the accompanying orthorhombic distortion.

\end{abstract}


\maketitle

The superconducting state of solid state systems has been one of the most highly investigated phenomena in condensed matter physics since its discovery 100 years ago \cite{onnes}. The newly unveiled layered Fe-based superconductors greatly restored the interest in the field. The first superconductivity report in this new system was for LaFeAsO with partial substitution of the oxygen site with fluorine \cite{LaFeAsOSC}. Soon latter, superconductivity with $T_c=38$ K was found in BaFe$_{2}$As$_{2}$ partially doped with potassium on the barium site \cite{RotterPRL}. In common, both undoped parent compounds presented a spin-density-wave (SDW) ordering transition connected with a tetragonal-to-orthorhombic structural phase transition \cite{LaFeAsO,RotterPRB}. For BaFe$_{2}$As$_{2}$ $T_{SDW}\approx138$ K vanishes continuously by K substitution on the Ba site \cite{RotterPRL} or Co \cite{Codoping}, Ni \cite{Nidoping}, Rh or Pd \cite{RhPddoping} substitution on the Fe site. The superconducting state emerges within a finite range of substitutions, resembling the high-$T_c$ cuprates phase diagram. Chemical substitutions also suppress and split the structural-magnetic transition \cite{RhPddoping,NMRurbano}. Isovalent substitutions such as P on the As site \cite{Pdoping} or Ru on the Fe site \cite{Rudoping} also tune BaFe$_{2}$As$_{2}$ into the superconducting state. In addition, superconductivity has been observed in undoped single crystals ($T_c\approx30$~K at $P\sim55$~kbar) \cite{SCPressureSelf,Yamakazi,Duncan}. Accessing the superconducting state without the need of charge carrier doping clearly differs this class of compounds from the cuprates.

The most widely investigated system is BaFe$_{2-x}$Co$_{x}$As$_{2}$, partly due to the availability of high-quality homogeneous single crystals. A convenient approximation to investigate the effects of Co-substitution on the electronic structure under density functional theory (DFT) has been the virtual crystal approach, in which the extra nuclear charge of Co in the Fe crystallographic site is averaged out without the need of a supercell \cite{Codoping}. Under this approximation, a shift of the chemical potential with increasing Co content is computed, similar to that predicted by electron doping. Angular resolved photoemission spectroscopy (ARPES) experiments show an evolution of the hole and electron pockets of the Fermi surface with $x$ that is consistent with this shift \cite{ChangLiu,Neupane}. In fact, the Co substituted system has been widely referred to as electron-doped BaFe$_2$As$_2$. This terminology brings an implicit assumption, still not verified experimentally by an element-specific probe, that Co substitution is able to tune the Fe electronic occupation. However, the real space density distribution of the extra $d$ electrons brought by Co substitution cannot be estimated under the virtual crystal approximation. Rather, a supercell approach to DFT with a distribution of 87.5\% of Fe and 12.5\% of Co has been applied, leading to the prediction that the excess $d$ electrons from the impurity are actually concentrated at the substitute Co site \cite{prliso} with little effect on the charge density distribution of the rest of the material (see also Ref. \onlinecite{prbiso}). Whether or not Co substitution is able to charge dope the Fe ions in this system is a major issue that may guide the identification of the mechanism of superconductivity in the Fe pnictides.

X-ray absorption near edge structure (XANES) spectroscopy is a classic probe to determine element-specific electronic ground states. In the electric dipolar absorption process involved in the Fe $K$ near edge, for instance, a photon-induced electronic transition from the Fe $1s$ core level to Fe $4p$ unoccupied states takes place. The energy of core level and end states are modified in distinct ways by local changes in electronic occupation due to the characteristic Coulomb interactions of each level with the doped electron or hole, causing a shift of the threshold absorption energy.
In this work we study the Fe $K$ near edge structure in high quality In-flux grown BaFe$_{2-x}$Co$_x$As$_{2}$ single crystals to determine whether the Co substitution is able to alter significantly the Fe electronic occupation.

BaFe$_{2-x}$Co$_x$As$_2$ single crystals were synthesized for $x=0.00$, 0.12, 0.17, 0.27 and 0.38 by the In flux method. Details of the growth process will be described elsewhere \cite{Influx}. The resulting single crystals showed typical linear dimensions in the $ab$ plane between approximately 0.5 and 2~mm. The actual Co:Fe ratios were determined by comparing the Fe and Co $K$ absorption edge steps and are found to be very close to the nominal values. In-plane electrical resistivity was measured using the standard four-probe method in a commercial physical properties measurement system (PPMS).
Room temperature XANES measurements were performed in the XAFS-2 beamline at the Brazilian Synchrotron Light Laboratory (LNLS). Crystals with thicknesses of a few microns along the $c$ direction were selected. Spectra were measured with 0.2~eV step width. At least two spectra were collected for each measured sample, in order to check for reproducibility of the spectral features and to improve the statistics. The edge-step normalization of the data was performed after a linear pre-edge subtraction and the regression of a quadratic polynomial beyond the edge, using the software ATHENA \cite{ATHENA}. The energy calibration was performed for each spectrum by simultaneously measuring and aligning the $K$ absorption edge of a standard Fe metal foil.
\textit{Ab initio} calculations for the XANES spectra were obtained using the FEFF8 code \cite{Ankudinov}, taking the reported tetragonal crystal structure of BaFe$_{2}$As$_{2}$ at 297~K (space group $I4/mmm$) \cite{RotterPRB} as the initial model. We adopt the Hedin-Lundqvist exchange potential with an imaginary part of 0.7~eV to account for the experimental broadening and a Debye-Waller factor of $\sigma^2=0.00465$~{\AA}$^2$ \cite{EXAFS_bafeas}. The atomic potential was calculated self consistently using a cluster of up to 32 atoms within a radius of 5.5~{\AA}. The full multiple scattering XANES calculations converges for a cluster of 154 atoms within the radius of 9.0~{\AA}. A rigid shift of 2.8~eV to lower energies was applied to all calculated XANES spectra to match with the experimental data. This shift has no influence on the relative changes of calculated spectral features under electron and hole doping that guide the main conclusions of this work.

\begin{figure}[htb!]
\begin{center}
\includegraphics[width=0.5\textwidth]{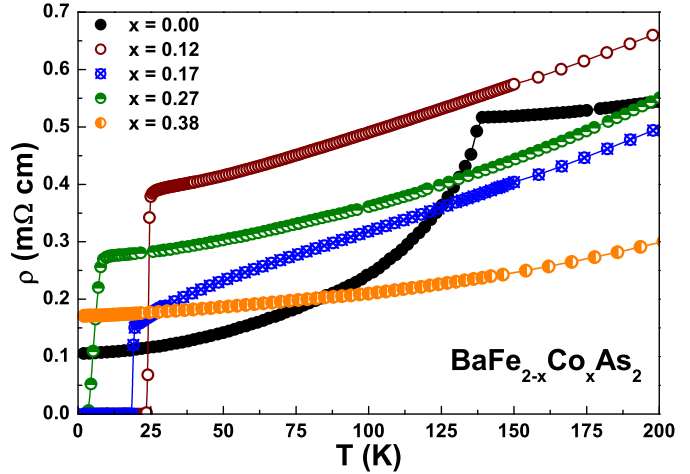}
\end{center}
\caption{\label{rho} (Color online) In-plane electrical resistivity of BaFe$_{2-x}$Co$_x$As$_2$ as a function of $T$. The solid lines are guides to the eyes.}
\end{figure}

Figure~\ref{rho} shows the electrical resistivity in the $ab$-plane for BaFe$_{2-x}$Co$_x$As$_2$ as a function of $T$. For $x=0.00$, the simultaneous structural and SDW transition is found to be around 138~K, in accordance with previous reports in self-flux single \cite{PRLFeAsflux} and polycrystalline \cite{RotterPRB} samples. With increasing $x$ we clearly see the suppression of the magnetic transition and the appearance of a superconducting state. The value of $T_c$ as a function of $x$ follows the previously reported phase diagrams, where a superconducting dome exists for $0.05\lesssim x\lesssim0.30$ for a optimal superconducting temperature at $x \approx 0.12$ \cite{co_phase1,co_phase2,co_phase3}.

\begin{figure}[htb!]
\begin{center}
\includegraphics[width=0.5\textwidth]{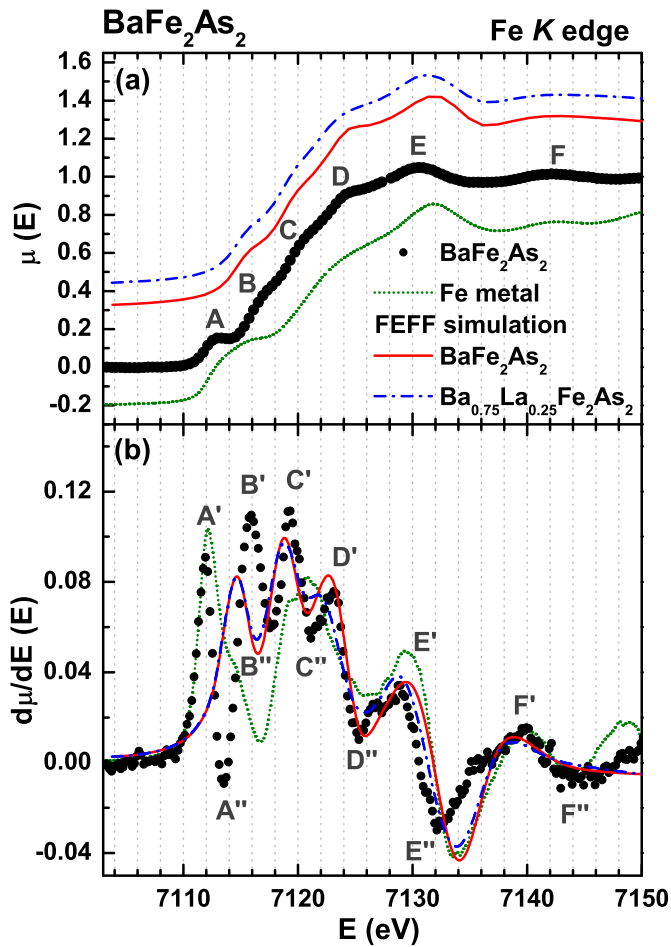}
\end{center}
\caption{\label{pure} (Color online) (a) Observed and calculated Fe $K$ edge XANES spectrum of pure BaFe$_2$As$_2$; calculated spectrum for the model compound Ba$_{0.75}$La$_{0.25}$Fe$_2$As$_2$ to simulate the effects of true electron doping. Prominent peaks and shoulders are indicated $A-F$. The calculated spectra are shifted vertically for better visualization. (b) Energy first derivative of the spectra shown in (a). Derivative maxima and minima are represented as $A'-F'$ and $A''-F''$, respectively. The reference Fe metal XANES spectrum is also shown for comparison.}
\end{figure}

The normalized Fe $K$ edge XANES spectrum $\mu(E)$ of BaFe$_2$As$_2$ is given in Fig.~\ref{pure}(a). Figure~\ref{pure}(b) shows the first derivative spectrum d$\mu(E)$/dE. Six distinct peaks or shoulders are noticed in the spectral region of interest and are labeled as $A-F$ in Fig.~\ref{pure}(a). To each peak or shoulder in this figure a corresponding maximum and a minimum are identified in the first derivative spectrum of Fig.~\ref{pure}(b) as $A'-F'$ and $A''-F''$, respectively. The simulated XANES spectrum of BaFe$_2$As$_2$ and its energy derivative are also shown in Fig.~\ref{pure} (solid line). The calculated spectrum captures fairly well the observed $B-F$ features. The $C-F$ features above the edge are dipolar transitions to unoccupied Fe $p$ projected states. An alternative simulation excluding $1s \rightarrow 3d$ quadrupolar transitions (not shown) shows a slightly weaker spectral weight for the $B$ shoulder, demonstrating that it originates partly from such transitions and partly from dipolar transitions allowed by $4p-3d$ mixing in the Fe site without inversion symmetry \cite{deGroot,Westre}. The observed $A$ pre-edge peak is completely absent in the simulation. This is possibly because charge-transfer effects in the absorption process, not fully taken into account in the simulation, pull down the $3d$ states yielding a combination of a well screened peak $B$ and a poorly screened peak $A$, as described in detail in Refs.~\onlinecite{deGroot} and \onlinecite{Bair}. The reference Fe metal XANES spectrum is also shown for comparison.

\begin{figure}[htb!]
\begin{center}
\includegraphics[width=0.5\textwidth]{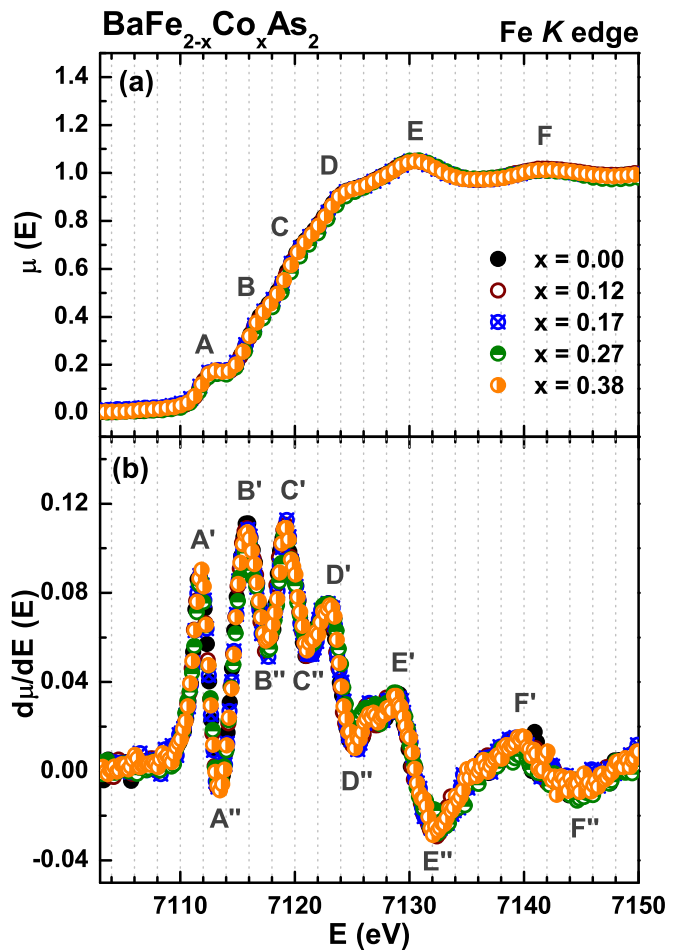}
\end{center}
\caption{\label{FeK} (Color online) (a) Normalized Fe $K$ edge XANES spectra of BaFe$_{2-x}$Co$_x$As$_2$ at room $T$. Prominent peaks and shoulders are indicated $A-F$. (b) First derivative of the XANES spectra in (a). Derivative maxima and minima are represented as $A'-F'$ and $A''-F''$, respectively.}
\end{figure}

\begin{figure}[htb!]
\begin{center}
\includegraphics[width=0.5\textwidth]{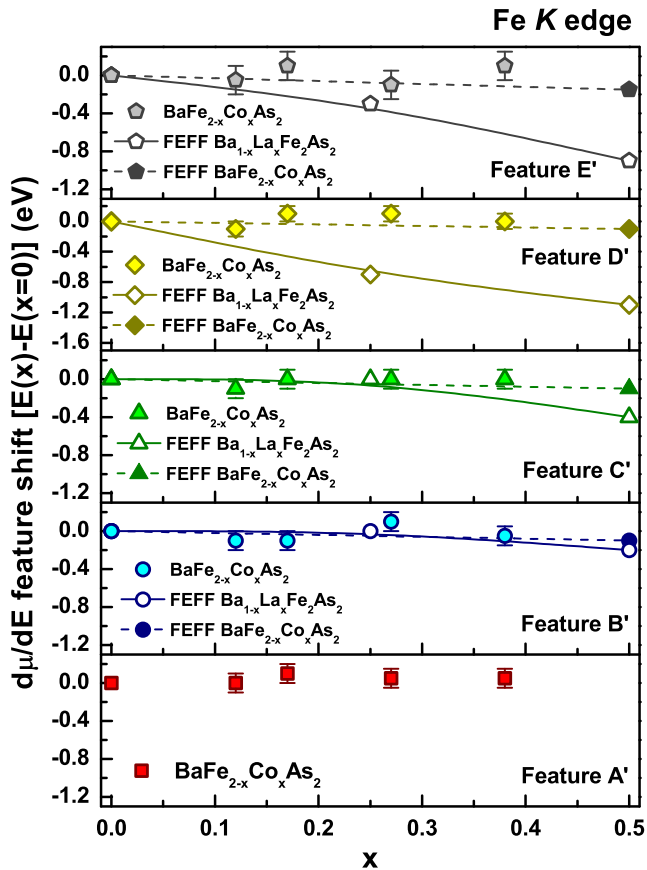}
\end{center}
\caption{(Color online) Fe $K$ edge XANES spectra first derivative $A'$, $B'$, $C'$, $D'$ and $E'$ feature (see Fig.~\ref{FeK}) position as a function of Co substitution in BaFe$_{2-x}$Co$_x$As$_2$. The solid spline lines are the expected red shifts by true electron doping, obtained by simulated XANES spectra of Ba$_{1-x}$La$_x$Fe$_2$As$_2$ ($x=0.00$, 0.25 and 0.50) model compounds. The dashed lines are the expected shifts obtained by a linear extrapolation using calculated XANES spectrum of BaFe$_{1.5}$Co$_{0.5}$As$_2$ (see text).} \label{positions2}
\end{figure}

Figures~\ref{FeK}(a) and \ref{FeK}(b) show the experimental Fe $K$ edge XANES spectra of BaFe$_{2-x}$Co$_x$As$_2$ and the corresponding energy-derivatives, respectively. Figure~\ref{positions2} shows the $x$ dependence of the positions of the $A'$, $B'$, $C'$, $D'$ and $E'$ features of the spectra. These results indicate no observable change in the Fe $K$ edge XANES spectra of BaFe$_2$As$_2$ under Co substitution.

Previous $K$ edge X-ray absorption spectroscopy studies on transition metal compounds with controlled valence such as La$_{1-x}$Ca$_x$MnO$_3$ showed edge shifts of a few eV per valence unit \cite{LaCaMnO}. Also, electron Nd$_{2-x}$Ce$_x$CuO$_{4-y}$ \cite{hightcXANESce} and hole La$_{2-x}$(Sr,K)$_x$CuO$_{4}$ \cite{hightcXANESce,hightcXANESsr,hightcXANESk} doped cuprate superconductors showed a Cu $K$ edge shift as a function of carrier concentration. For the cuprates it is well established that charge carrier doping is fundamental for the appearance of superconductivity and XANES can indeed probe changes of the local electronic structure of the absorbing atom. In general, red (blue) shifts are observed for electron (hole) doping. The lack of an Fe $K$ edge absorption threshold energy shift in Co substituted BaFe$_{2}$As$_{2}$ implies that Co is not charge doping the Fe ions which are thought to be responsible for the superconductivity.

In order to quantify the expected effects of true Fe electron doping on the XANES spectrum of BaFe$_2$As$_2$, the relative shifts of the simulated $B'-E'$ features in Fig. \ref{pure}(b) was computed for the model system Ba$_{1-x}$La$_{x}$Fe$_2$As$_2$. To account for the La substitution within the cluster of atoms in FEFF, Ba ions were randomly replaced by La while keeping the desired stoichiometry in each neighboring layer. The same atomic positions of the pure compound were employed. The simulated spectrum for $x=0.25$ is given in Figs.~\ref{pure}(a) and \ref{pure}(b). The calculated energy shift of these features as a function of $x$ are displayed in Fig.~\ref{positions2} as empty symbols and solid spline lines. These shifts are obtained relative to the Fe $K$ edge XANES simulation for pure BaFe$_2$As$_2$. It is clear that a red shift of some of the features of the XANES spectrum is computed under electron doping, most notably the $D'$ and $E'$ features. As already mentioned, such red shifts under electron doping are largely expected based on previous $K$ edge XANES experiments on other transition metal compounds \cite{LaCaMnO,hightcXANESce,hightcXANESsr,hightcXANESk}. We should mention that the calculated positions of the $B'$ and  $C'$ show much less influence of electron doping, indicating a non-rigid shift of the Fe $K$ edge under doping. In any case, the contrast between the observed constant positions of the $D'$ and $E'$ features under Co substitution and the calculated red shifts of up to $\sim 1$~eV for $x=0.50$ electron doping (see Fig. \ref{positions2}) unambiguously demonstrates that Co substitution does not induce Fe valence changes in BaFe$_2$As$_2$. We have also preformed FEFF calculation for BaFe$_{2-x}$Co$_x$As$_2$ ($x=0.5$). In this calculation, the extra Co electron is found to stay entirely within the muffin-tin sphere of the substitute atom, consistent with Ref. \onlinecite{prliso}. The closed symbols and dashed lines in Fig. \ref{positions2} show the computed shift for $x=0.5$ and a linear extrapolation for $0 < x < 0.5$, respectively. No significant shifts of the Fe K edge features are calculated, in agreement with our experimental results.

Co substitution brings an extra electron to the transition-metal $3d$ bands close to the Fermi level and has an important impact in the Fermi surface, consistent with a shift of the chemical potential by Co substitution in first approximation \cite{Codoping,ChangLiu,Neupane,RudopingARPES}. This fact is complementary rather than contradictory with the constant Fe electronic occupation predicted in Ref. \onlinecite{prliso} and demonstrated here. In fact, the electronic states close to the Fermi level show mixed Fe and Co $3d$ character, therefore a tuning of the chemical potential may occur without necessarily changing the total electronic occupation of Fe.

The rigid Fe valence under Co substitution raises the question on the nature of the perturbation that weakens the SDW ground state allowing the emergence of superconductivity in BaFe$_{2-x}$Co$_x$As$_2$ and other related systems. Since the SDW transition is coupled with a significant orthorhombic lattice distortion, the extra elastic energy related with the SDW order can only be overcome by the gain in exchange energy if the local spins are above a threshold value. The local Fe $3d$ spins are influenced by hybridization with electronic states of neighboring atoms, which may be tuned either by external pressure or by atomic substitutions. In fact, despite not provoking a significant local electronic charge instability in the Fe ions, Co substitution does alter the local atomic structure, and a slight decrease of the As-(Fe,Co) bond length has been observed \cite{EXAFS_bafeas}, which is possibly connected with a reduction of the Fe local moments \cite{Johannes}. Also, the Co ions have enhanced hybridization with As with respect to Fe \cite{Codoping}, which may further increase the elastic energy penalty of the orthorhombic distortion associated with the SDW order and help destabilizing this ground state with respect to the superconducting one.

In summary, our XANES measurements in BaFe$_{2-x}$Co$_x$As$_{2}$ single crystals reveal that the Fe $K$ absorption edge is unaltered by Co substitution. This result is compatible with a rigid Fe valence. Superconductivity in this system may emerge due to Fe bonding modifications induced by the substitute atom that destabilizes the SDW ground state.

\begin{acknowledgments}
This work was supported by FAPESP and CNPq (Brazil). LNLS is acknowledged for concession of beamtime.
\end{acknowledgments}

\end{document}